\documentclass[conference]{IEEEtran}
\usepackage{amsmath}
\usepackage{amssymb}
\usepackage{arxivmydefins}
\usepackage[dvipdfmx]{graphicx}
\usepackage{layout}
\def\QED{\hfill$\Box$}
\definecolor{Bgreen}{rgb}{ .0, .55, .0 }
\definecolor{Red}{rgb}{ 1.0, .0, .0 }
\definecolor{Navy}{rgb}{ 0.0, .0, 1.0 }

\begin{document}
%
\title{Channel Resolvability Theorems \\ for General Sources and Channels}

\author{
 \IEEEauthorblockN{Hideki Yagi}
  \IEEEauthorblockA{
Dept.\ of Computer and Network Engineering\\
University of Electro-Communications \\
Tokyo, Japan\\
Email: h.yagi@uec.ac.jp}
}

%

\newcommand{\FLS}{S}
\newcommand{\OFLS}{S^*}
\newcommand{\Vn}{V^n}
\newcommand{\vectV}{\vect{V}}
\newcommand{\Zn}{Z^n}
\newcommand{\vectZ}{\vect{Z}}
\newcommand{\awidehat}{\hat}


\maketitle

\begin{abstract}
In the problem of channel resolvability, where a given output probability distribution via a channel is approximated by transforming the uniform random numbers, characterizing the asymptotically minimum rate of the size of the random numbers, called the channel resolvability, has been open.
This paper derives formulas for the channel resolvability for a given general source and  channel pair.
We also investigate the channel resolvability in an optimistic sense.
It is demonstrated that the derived general formulas recapture a single-letter formula for the stationary memoryless source and channel.
When the channel is the identity mapping, the established formulas reduce to an alternative form of the spectral sup-entropy rates, which play a key role in information spectrum methods.
The analysis is also extended to the second-order channel resolvability.
\end{abstract}


\renewcommand\thefootnote{}
\footnotetext{This research is supported by JSPS KAKENHI Grant Number JP16K06340.}
\renewcommand\thefootnote{\arabic{footnote}}

%
\IEEEpeerreviewmaketitle

\section{Introduction}

Finding the asymptotically minimum rate of the size of the uniform random numbers (\emph{channel resolvability}) which can approximate a given target output distribution via a channel is called the problem of \emph{channel resolvability}.
When  the variational distance between the target output distribution and the approximated distribution is required to be asymptotically not greater than $\delta \in [0,1)$, the problem is called the problem of $\delta$-channel resolvability.
Though these problems were introduced by Han and Verd\'u \cite{Han-Verdu1993} more than two decades ago, the general formula for the channel resolvability has not been known in general.
A few cases where the channel resolvability has been characterized are the worst input case with $\delta =0$ by Hayashi \cite{Hayashi2006} and the case of the stationary memoryless source and channel by Watanabe and Hayashi \cite{Watanabe-Hayashi2014}.
Recently, much attention has been paid to the channel resolvability because this technique can be used to guarantee the strong secrecy in physical-layer security systems \cite{Bloch-Laneman2013,Hayashi2006}.
Thus, it is desirable to characterize the channel resolvability for a given pair of the input distribution and the general channel. 

In this paper, we characterize the $\delta$-channel resolvability for a general source and a general channel with any $\delta \in [0,1)$. By taking the maximum over all possible general sources, we can naturally obtain the general formula for the worst input case.
We also investigate the  $\delta$-channel resolvability in an optimistic sense.
When we restrict ourselves to the noiseless channel (identity mapping), the problem of channel resolvability reduces to the problem of \emph{source resolvability} \cite{Han-Verdu1993,Steinberg-Verdu1996}. 
The established general formula provides a new expression for the $\delta$-spectral sup-entropy rate, which is a well-known information quantity in information spectrum methods \cite{Han2003}.
The analysis is also extended to the \emph{second-order} channel resolvability, which is defined as the asymptotically minimum second-order rate of the size of uniform random numbers with respect to a fixed first-order resolvability rate.

\section{Problem Formulation: Channel Resolvability}

Let $\mathcal{X}$ and $\mathcal{Y}$ be finite or countably infinite alphabets.
Let $X^n$ denote a sequence of $n$ random variables taking values in $\mathcal{X}^n$ with probability distribution $P_{X^n}$.
In this paper,  we identify $P_{X^n}$ with $X^n$, and both expressions are used interchangeably.
We call $\vect{X} = \{ X^n\}_{n =1}^\infty$ a \emph{general source}. 
Also, let  $W^n : \mathcal{X}^n \rightarrow \mathcal{Y}^n$ denote a stochastic mapping, and we call $\vect{W} = \{W^n \}_{n=1}^\infty$ a \emph{general channel}.
We do not impose any assumptions such as stationarity or ergodicity on either $\vect{X}$ or $\vect{W}$.
We denote by $\vect{Y} = \{ Y^n\}_{n =1}^\infty$ the output process via $\vect{W}$ due to input process $\vect{X}$.

\begin{figure}[t]
\begin{center}
\includegraphics[width=8.0cm]{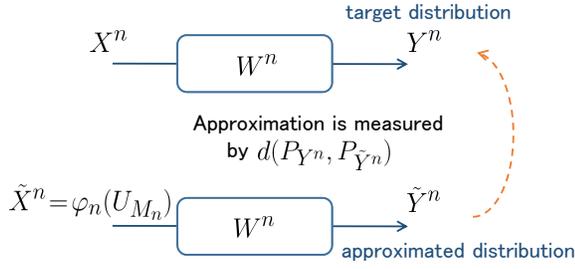}
\end{center}
\caption{Channel Resolvability System} \label{fig:ch_resolvability}
\end{figure}

\smallskip
We review the problem of \emph{channel resolvability} \cite{Han2003} using the variational distance as an approximation measure.
Let ${U_{M_n}}$ denote the \emph{uniform random number} of size $M_n$, which is a random variable \emph{uniformly} distributed over $\{ 1, \ldots, M_n\}$.
Consider approximating the \emph{target distribution} $P_{Y^n}$ by using ${U_{M_n}}$ via a deterministic mapping $\varphi_n : \{ 1, \ldots, M_n\} \rightarrow \mathcal{X}^n$ and $W^n$.
We denote 
by $P_{\tilde{Y}^n}$ the approximated output distribution via $W^n$ due to the input $\tilde{X}^n := \varphi_n({U_{M_n}})$  (cf.\ Fig.\ \ref{fig:ch_resolvability}).
Precision of the approximation is measured by the \emph{variational distance} between $P_{Y^n}$ and $P_{\tilde{Y}^n} $.
\begin{e_defin}[Variational Distance] \label{def:variational_dist}
{\rm
~~~Letting $P_{Z}$ and $P_{\tilde{Z}}$ be probability distributions on a countably infinite set $\mathcal{Z}$,
 \begin{align}
 d(P_{Z}, P_{\tilde{Z}}) := \frac{1}{2} \sum_{z\in \mathcal{Z}} |P_{Z}(z) - P_{\tilde{Z}} (z)|  
 \end{align}
is called the \emph{variational distance} between $P_{Z}$ and $P_{\tilde{Z}}$.}
\QED
\end{e_defin}
It is easily seen that $0 \le  d(P_{Z}, P_{\tilde{Z}}) \le 1$, where the left inequality becomes equality if and only if $P_{Z} = P_{\tilde{Z}}$.

For any given sequence of random variables $\{Z_n\}_{n=1}^\infty$, we introduce quantities which play an important role in \emph{information spectrum methods} \cite{Han2003}.
 \begin{e_defin}[$\varepsilon$-Limit Superior in Probability]~~For $\varepsilon \in [0,1]$,
\begin{align}\vspace*{-3mm}
\hspace*{-3mm} \mbox{\rm $\varepsilon$p-}\limsup_{n \rightarrow \infty} Z_n & \! := \!  \inf \left\{ \alpha : \limsup_{n \rightarrow \infty} \Pr \{Z_n \!   > \! \alpha \} \! \le \! \varepsilon \right\}, \label{eq:quantity1} \\
\hspace*{-3mm}\mbox{\rm $\varepsilon$p$^*$-}\limsup_{n \rightarrow \infty} Z_n &\! := \!  \inf \left\{ \alpha  : \liminf_{n \rightarrow \infty} \Pr \{Z_n > \alpha \} \! \le \! \varepsilon  \right\}.   \label{eq:quantity2} 
\end{align}
For $\varepsilon=0$, the right-hand sides of \eqref{eq:quantity1} and \eqref{eq:quantity2} are simply denoted by $\displaystyle \mbox{\rm p-}\limsup_{n \rightarrow \infty} Z_n$ and $\displaystyle \mbox{\rm p$^*$-}\limsup_{n \rightarrow \infty} Z_n$, respectively.
\QED
\end{e_defin}

\medskip

The problem of channel resolvability has been introduced by Han and Verd\'u \cite{Han-Verdu1993}. 

\begin{e_defin}[$\delta$-Channel Resolvability]
{\rm
Let $\delta \in [0,1)$ be fixed arbitrarily.
A resolvability rate $R \ge 0$ is said to be \emph{$\delta$-achievable at $\vect{X}$} if there exists a deterministic mapping $\varphi_n : \{ 1, \ldots, M_n\} \rightarrow \mathcal{X}^n$ satisfying
\begin{align}
 \limsup_{n \rightarrow \infty} \frac{1}{n} \log M_n &\le R, \label{eq:rate_cond}  \\ 
 \limsup_{n \rightarrow \infty} d(P_{Y^n}, P_{\tilde{Y}^n}) &\le \delta \label{eq:variational_dist_cond},
\end{align}
where $\tilde{Y}^n$ denotes the output via $W^n$ due to the input $\tilde{X}^n = \varphi_n({U_{M_n}})$.
We define
\begin{align}
\FLS (\delta | \vect{X}, \vect{W}):= \inf \{ R : ~ R ~\mbox{is $\delta$-achievable at}~\vect{X} \}, \label{eq:fixed_length_rate}
\end{align}
which is called the \emph{$\delta$-channel resolvability} (at $\vect{X})$.}
\QED
\end{e_defin}

Equation \eqref{eq:variational_dist_cond} requires $d(P_{Y^n}, P_{\tilde{Y}^n}) \le \delta + \gamma$ for all large $n$, where $\gamma >0$ is an arbitrary constant.
We may consider a slightly weaker constraint, which requires $d(P_{Y^n}, P_{\tilde{Y}^n}) \le \delta + \gamma$ for infinitely many $n$
The following problem is the weaker version of the $\delta$-channel resolvability, introduced by \cite{Steinberg1998} in the context of partial resolvability.
\begin{e_defin}[Optimistic $\delta$-Channel Resolvability]
{\rm
Let $\delta \in [0,1)$ be fixed arbitrarily.
A resolvability rate $R \ge 0$ is said to be \emph{optimistically $\delta$-achievable at $\vect{X}$} if there exists a deterministic mapping $\varphi_n : \{ 1, \ldots, M_n\} \rightarrow \mathcal{X}^n$ satisfying
\begin{align}
 \limsup_{n \rightarrow \infty} \frac{1}{n} \log M_n &\le R, \label{eq:opt_rate_cond}  \\ 
 \liminf_{n \rightarrow \infty} d(P_{Y^n}, P_{\tilde{Y}^n}) &\le \delta \label{eq:opt_variational_dist_cond}.
\end{align}
We define
\begin{align}
&\OFLS (\delta | \vect{X}, \vect{W}) \nonumber \\
&~:= \inf \{ R : ~ R ~\mbox{is optimistically $\delta$-achievable at}~\vect{X} \}, \nonumber 
\end{align}
referred to as the \emph{optimistic $\delta$-channel resolvability} (at $\vect{X})$.
}\QED
\end{e_defin}

\medskip

The following channel resolvability theorem is implicitly proved by Hayashi \cite{Hayashi2006} for general sources and channels.
\begin{e_theo}[{Hayashi \cite{Hayashi2006}}] \label{theo:fixed_length_resolvability}
Let $\delta \in [0,1)$ be fixed arbitrarily.
For any general source $\vect{X} = \{ X^n\}_{n=1}^\infty $ and any general channel $\vect{W}  = \{ W^n\}_{n=1}^\infty$,
\begin{align}
\FLS (\delta | \vect{X}, \vect{W}) &\le \overline{I}_\delta (\vect{X};\vect{Y}), \label{eq:fixed_length_achievable} \\
\OFLS (\delta | \vect{X}, \vect{W}) &\le \overline{I}^*_\delta (\vect{X};\vect{Y}),  \label{eq:opt_fixed_length_achievable} 
\end{align}
where we define
\begin{align}
&\overline{I}_\delta(\vect{X}; \vect{Y}) := \mbox{\rm $\delta$p-}\limsup_{n \rightarrow \infty}  \frac{1}{n} \log \frac{W^n(Y^n | X^n)}{P_{Y^n}(Y^n)},\\
&\overline{I}_\delta^*(\vect{X}; \vect{Y}) := \mbox{\rm $\delta$p$^*$-}\limsup_{n \rightarrow \infty}  \frac{1}{n} \log \frac{W^n(Y^n | X^n)}{P_{Y^n}(Y^n)}. 
\end{align}
\QED
\end{e_theo}

Unfortunately, Theorem \ref{theo:fixed_length_resolvability} does not provide a lower bound on the $\delta$-channel resolvability.
For the worst input case, in contrast, a lower bound has also been given by Hayashi \cite{Hayashi2006}.
\begin{e_theo}[{Hayashi \cite{Hayashi2006}}] \label{theo:fixed_length_resolvability2}
{\rm
For any general channel $\vect{W}  = \{ W^n\}_{n=1}^\infty$,
\begin{align}
\sup_{\vect{X}} \overline{I}_{2\delta}(\vect{X};\vect{Y}) \le \sup_{\vect{X}} \FLS (\delta | \vect{X}, \vect{W}) \le \sup_{\vect{X}} \overline{I}_{\delta}(\vect{X};\vect{Y}), \label{eq:fixed_length_capacity_bound1}  \\
\sup_{\vect{X}} \overline{I}_{2\delta}^* (\vect{X};\vect{Y}) \le \sup_{\vect{X}} \OFLS (\delta | \vect{X}, \vect{W}) \le \sup_{\vect{X}} \overline{I}_{\delta}^* (\vect{X};\vect{Y}). \label{eq:fixed_length_capacity_bound2} 
\end{align}
In particular, 
\begin{align}
\sup_{\vect{X}}  \FLS (0 | \vect{X}, \vect{W}) &= \sup_{\vect{X}} \overline{I}(\vect{X};\vect{Y}), \\
\sup_{\vect{X}}  \OFLS (0 | \vect{X}, \vect{W}) &= \sup_{\vect{X}} \overline{I}^*(\vect{X};\vect{Y}), \label{eq:0-fixed_length_capacity_formula} 
\end{align}
where we define
\begin{align}
&\overline{I}(\vect{X}; \vect{Y}) := \mbox{\rm p-}\limsup_{n \rightarrow \infty}  \frac{1}{n} \log \frac{W^n(Y^n | X^n)}{P_{Y^n}(Y^n)},\\
&\overline{I}^*(\vect{X}; \vect{Y}) := \mbox{\rm p$^*$-}\limsup_{n \rightarrow \infty}  \frac{1}{n} \log \frac{W^n(Y^n | X^n)}{P_{Y^n}(Y^n)}. 
\end{align}
}\QED
\end{e_theo}

\section{Main Theorems: $\delta$-Channel Resolvability}

Now, we give the general formulas for the $\delta$-channel resolvability at a specific input $\vect{X}$ and its optimistic version.
\begin{e_theo}\label{theo:d-fixed-length_resolvability}
Let  $\delta \in [0,1)$ be fixed arbitrarily.
For any input process $\vect{X}$ and any general channel $\vect{W}$, 
\begin{align}
\FLS (\delta | \vect{X}, \vect{W}) 
&=\inf_{\hat{\vect{X}} \in B_\delta (\vect{X}, \vect{W})} \overline{I} (\hat{\vect{X}}; \hat{\vect{Y}}), \label{eq:d-fixed-length_general_formula1} \\
\OFLS (\delta | \vect{X}, \vect{W}) 
&=\inf_{\hat{\vect{X}} \in B_\delta^* (\vect{X}, \vect{W})} \overline{I} (\hat{\vect{X}}; \hat{\vect{Y}}), \label{eq:opt_d-length_general_formula1} 
\end{align}
where $\hat{\vect{Y}} = \{ \awidehat{Y}^n\}_{n=1}^\infty$ denotes the output process via $\vect{W}$ due to the input process $\hat{\vect{X}} = \{ \awidehat{X}^n\}_{n=1}^\infty$, and we define
\begin{align}
B_\delta (\vect{X}, \vect{W}) &:= \left\{ \hat{\vect{X}} = \{ \awidehat{X}^n\}_{n=1}^\infty: \limsup_{n \rightarrow \infty} d(P_{Y^n}, P_{\awidehat{Y}^n}) \le \delta \right\}, \nonumber \\ 
B_\delta^* (\vect{X}, \! \vect{W}) &:=  \left\{ \hat{\vect{X}} \! = \! \big\{ \awidehat{X}^n\big\}_{n=1}^\infty \! :  \liminf_{n \rightarrow \infty} d(P_{Y^n}, P_{\awidehat{Y}^n}) \! \le \! \delta \right\}.  \nonumber
\end{align}
\end{e_theo}
\begin{e_proof}
The proof is given in Sec.\ \ref{sect:proof_theorems}.
\QED
\end{e_proof}
\begin{e_rema} \label{rema:right_continuity}
The right-hand sides of \eqref{eq:d-fixed-length_general_formula1} and \eqref{eq:opt_d-length_general_formula1} are nonincreasing functions of $\delta$.
Furthermore, these are right-continuous in $\delta \in [0,1)$.
\QED
\end{e_rema}

\begin{e_rema}  \label{rema:comparison_general_formulas}
As is mentioned in Theorem \ref{theo:fixed_length_resolvability}, Hayashi \cite[Theorem 4]{Hayashi2006} has implicitly shown that any rate $R > \overline{I}_{\delta}(\vect{X};\vect{Y})$ is $\delta$-achievable at a specific input $\vect{X}$.
Therefore, we obtain the following relation between the right-hand side of \eqref{eq:d-fixed-length_general_formula1} and $\delta$-spectral sup-mutual information rate $\overline{I}_{\delta}(\vect{X};\vect{Y})$:
\begin{align}
\inf_{\hat{\vect{X}} \in B_\delta (\vect{X}, \vect{W})} \overline{I} (\hat{\vect{X}}; \hat{\vect{Y}}) \le \overline{I}_{\delta}(\vect{X};\vect{Y})~~~(\delta \in [0,1)) \label{eq:general_formula_relation}
\end{align}
and analogously
\begin{align}
\inf_{\hat{\vect{X}} \in B_\delta^* (\vect{X}, \vect{W})} \overline{I} (\hat{\vect{X}}; \hat{\vect{Y}}) \le \overline{I}_{\delta}^*(\vect{X};\vect{Y})~~~(\delta \in [0,1)). \label{eq:opt_general_formula_relation}
\end{align}
We can find examples of $\vect{X}$ and $\vect{W}$ for which the inequalities in \eqref{eq:general_formula_relation} and \eqref{eq:opt_general_formula_relation} are strict.
This statement is also true even in the case $\delta = 0$.
\QED
\end{e_rema}

\medskip
Although the formulas established in Theorem \ref{theo:d-fixed-length_resolvability} are sufficient to characterize $\FLS (\delta | \vect{X}, \vect{W})$ and $\OFLS (\delta | \vect{X}, \vect{W})$,
it requires a tedious task to derive a single-letter formula for the stationary memoryless source and channel pair.  
We give alternative formulas in the following theorem:
\begin{e_theo}\label{theo:alt_fixed-length_resolvability}
Let  $\delta \in [0,1)$ be fixed arbitrarily.
For any input process $\vect{X}$ and any general channel $\vect{W}$, 
\begin{align}
\FLS (\delta | \vect{X}, \vect{W}) &=\inf_{\hat{\vect{X}} \in B_\delta (\vect{X}, \vect{W})} \sup_{\substack{\varepsilon \ge 0, \\ \vect{Z} \in \tilde{B}_{\varepsilon} (\hat{\vect{Y}})}}   \overline{D}_{\varepsilon} (\vect{W}|| \vect{Z} | \hat{\vect{X}}), \label{eq:d-fixed-length_general_formula2} \\
\OFLS (\delta | \vect{X}, \vect{W}) &=\inf_{\hat{\vect{X}} \in B_\delta^* (\vect{X}, \vect{W})} \sup_{\substack{\varepsilon \ge 0, \\ \vect{Z} \in \tilde{B}_{\varepsilon} (\hat{\vect{Y}})}}   \overline{D}_{\varepsilon} (\vect{W}|| \vect{Z} | \hat{\vect{X}}), \label{eq:opt_d-length_general_formula2} 
\end{align}
where $\hat{\vect{Y}} = \{ \awidehat{Y}^n\}_{n=1}^\infty$ denotes the output process via $\vect{W}$ due to input process $\hat{\vect{X}} = \{ \awidehat{X}^n\}_{n=1}^\infty$, and we define
\begin{align}
&\overline{D}_\varepsilon (\vect{W}|| \vect{Z}| \hat{\vect{X}}) :=  \mbox{\rm $\varepsilon$p-}\limsup_{n \rightarrow \infty} \frac{1}{n} \log \frac{W^n(\awidehat{Y}^n|\awidehat{X}^n)}{P_{Z^n}(\awidehat{Y}^n)}, \nonumber \\
&\tilde{B}_\varepsilon (\vect{Y}) := \left\{ \vect{Z} =\left\{ Z^n \right\}_{n=1}^\infty : \,  \limsup_{n \rightarrow \infty} d(P_{Y^n}, P_{Z^n}) \le \varepsilon \right\}. \nonumber 
\end{align}
\end{e_theo}
\begin{e_proof}
The proof is given in Sec.\ \ref{sect:proof_theorems}.
\QED
\end{e_proof}

\begin{e_rema} \label{rema:comparison_general_formulas2}
Theorems \ref{theo:d-fixed-length_resolvability} and \ref{theo:alt_fixed-length_resolvability} provide two formulas for the $\delta$-channel resolvability $S(\delta|\vect{X}, \vect{W})$.
Although the characterization in \eqref{eq:d-fixed-length_general_formula2} is more complicated, this expression can be seen as a counterpart of the alternative formula for the channel capacity given by Hayashi and Nagaoka \cite[Theorem 1]{Hayashi-Nagaoka2003} established for quantum channels. The corresponding formula for the $\delta$-channel capacity over classical channels can be found in \cite[Theorem 6]{Hayashi2009}.  
Comparing the two characterizations, the following inequality is obvious for all $\delta \in [0,1)$:
\begin{align}
\inf_{\hat{\vect{X}} \in B_\delta (\vect{X}, \vect{W})} \overline{I} (\hat{\vect{X}}; \hat{\vect{Y}}) &\le \inf_{\hat{\vect{X}} \in B_\delta (\vect{X}, \vect{W})} \sup_{\substack{\varepsilon \ge 0, \\ \vect{Z} \in \tilde{B}_{\varepsilon} (\hat{\vect{Y}})}}   \overline{D}_{\varepsilon} (\vect{W}|| \vect{Z} | \hat{\vect{X}}). \label{eq:general_formula_relation2}
\end{align}
because $\overline{D}_{0} (\vect{W}||  \hat{\vect{Y}} | \hat{\vect{X}}) = \overline{I} (\hat{\vect{X}}; \hat{\vect{Y}})$.
Also, we have
 for all $\delta \in [0,1)$:
\begin{align}
\inf_{\hat{\vect{X}} \in B_\delta^* (\vect{X}, \vect{W})} \overline{I} (\hat{\vect{X}}; \hat{\vect{Y}}) &\le \inf_{\hat{\vect{X}} \in B_\delta^* (\vect{X}, \vect{W})} \sup_{\substack{\varepsilon \ge 0, \\ \vect{Z} \in \tilde{B}_{\varepsilon} (\hat{\vect{Y}})}}   \overline{D}_{\varepsilon} (\vect{W}|| \vect{Z} | \hat{\vect{X}}). \label{eq:opt_general_formula_relation2}
\end{align}
These relationships are of use to prove Theorems \ref{theo:d-fixed-length_resolvability} and \ref{theo:alt_fixed-length_resolvability}. 
\QED
\end{e_rema}

\section{Proof of Theorems \ref{theo:d-fixed-length_resolvability} and \ref{theo:alt_fixed-length_resolvability}} \label{sect:proof_theorems}

\subsection{Finite-Length Bounds}
As we take an information spectrum approach to prove the general formulas in  Theorems \ref{theo:d-fixed-length_resolvability} and \ref{theo:alt_fixed-length_resolvability}, we will use finite-length upper and lower bounds on the variational distance,  which hold for each blocklength $n$.

In the proof of the direct part, we use the following lemma.
\begin{e_lem}[Finite-Length Upper Bound \cite{Hayashi2006}] \label{lem:achievability_bound}
{\rm
Let $V^n$ be an arbitrary input random variable, and its corresponding output via $W^n$ is denoted by $Z^n$.
Then, for any given positive integer $M_n$, there exists a mapping $\varphi_n: \{ 1, 2, \ldots, M_n\} \rightarrow \mathcal{X}^n$ such that
 \begin{align} \label{eq:achievability_bound}
 &  d(P_{Z^n}, P_{\tilde{Y}^n}) \nonumber \\
 &~\le  \Pr \left\{ \frac{1}{n} \log \frac{W^n(Z^n | V^n)}{P_{Z^n}(Z^n)} > c \right\} + \frac{1}{2}\sqrt{\frac{e^{nc}}{M_n}} , 
  \end{align} 
  where $c \ge 0$ is an arbitrary constant and $\tilde{Y}^n$ denotes the output via $W^n$ due to input $\tilde{X}^n = \varphi_n({U_{M_n}})$. 
}\QED
\end{e_lem}

In the proof of the converse part, we use the following lemma.
\begin{e_lem}[Finite-Length Lower Bound] \label{lem:FL-converse}
Let $P_{Z^n}$ be an arbitrary probability distribution on $\mathcal{Y}^n$. Then, for any uniform random number $U_{M_n}$ of size $M_n$ and a deterministic mapping  $\varphi_n : \{ 1, 2,\ldots,M_n\} \rightarrow \mathcal{X}^n$ we have
\begin{align}
\hspace*{-1mm} d(P_{Z^n}, P_{\tilde{Y}^n}) \ge \Pr  \left\{ \frac{1}{n} \log \frac{W^n(\tilde{Y}^n | \tilde{X}^n)}{P_{Z^n}(\tilde{Y}^n)} \ge c \right\} - \frac{M_n}{e^{nc}}, \label{eq:FL-converse}
\end{align}
where $\tilde{X}^n= \varphi_n(U_{M_n})$, $\tilde{Y}^n$ denotes the output via $W^n$ due to $\tilde{X}^n$, and $c $ is an arbitrary constant satisfying $M_n \le e^{nc}$. 
\end{e_lem}
\noindent
(\emph{Proof})~~
First, we define
\begin{align}
T_n := \left\{ \vect{y} \in \mathcal{Y}^n : P_{\tilde{Y}^n}(\vect{y}) \ge \frac{e^{nc}}{M_n} P_{Z^n}(\vect{y})\right\}.
\end{align}
Then, by the definition of the variational distance, it is easily verified that
\begin{align}
d(P_{Z^n}, P_{\tilde{Y}^n}) \ge P_{\tilde{Y}^n}(T_n) - P_{Z^n}(T_n), \label{eq:initial_LB}
\end{align}
where the second term on the right-hand side can be evaluated as
\begin{align}
 \hspace*{-2mm} P_{Z^n}(T_n) = \sum_{\vect{y} \in T_n} P_{Z^n}(\vect{y}) \le \frac{M_n}{e^{nc}} \sum_{\vect{y} \in T_n} P_{\tilde{Y}^n} (\vect{y}) \le \frac{M_n}{e^{nc}}. \label{eq:2nd_term}
\end{align}
To evaluate the first term on the right-hand side of \eqref{eq:initial_LB}, we borrow an idea given in \cite{Watanabe-Hayashi2014}.
Since 
\begin{align}
P_{\tilde{Y}^n}(\vect{y}) = \sum_{i=1}^{M_n} \frac{1}{M_n} W^n(\vect{y} | \varphi_n(i)) ~~~~(\vect{y} \in \mathcal{Y}^n),
\end{align}
denoting $ W^n_{\varphi_n(i)} (\vect{y}) = W^n(\vect{y}|\varphi_n(i))$, 
we have
\begin{align}
&P_{\tilde{Y}^n}(T_n) \nonumber \\
&~= \sum_{i=1}^{M_n} \frac{1}{M_n} W^n_{\varphi_n(i)} \left\{ P_{\tilde{Y}^n}(\tilde{Y}^n) \ge \frac{e^{nc}}{M_n} P_{Z^n}(\tilde{Y}^n)\right\}  \nonumber \\
&~  =  \sum_{i=1}^{M_n} \frac{1}{M_n} W^n_{\varphi_n(i)}\left\{ \sum_{j=1}^{M_n} \frac{1}{M_n} W^n_{\varphi_n(j)} (\tilde{Y}^n) \ge \frac{e^{nc}}{M_n} P_{Z^n}(\tilde{Y}^n)\right\}. \nonumber
 \end{align}
Here, noticing that
\begin{align}
&\frac{1}{M_n} W^n_{\varphi_n(i)} (\vect{y}) \ge  e^{nc} P_{Z^n}(\vect{y}) \nonumber \\
&~~\Longrightarrow ~~\sum_{j}\frac{1}{M_n} W^n_{\varphi_n(j)}(\vect{y}) \ge  e^{nc} P_{Z^n}(\vect{y}),
\end{align}
we obtain the following lower bound:
\begin{align}
P_{\tilde{Y}^n}(T_n) &\ge  \sum_{i=1}^{M_n} \frac{1}{M_n} W^n_{\varphi_n(i)}\left\{  W^n_{\varphi_n(i)} (\tilde{Y}^n) \ge e^{nc} P_{Z^n}(\tilde{Y}^n)\right\}.  \label{eq:1st_term}
\end{align}
Thus, plugging \eqref{eq:2nd_term} and \eqref{eq:1st_term} into \eqref{eq:initial_LB}, we obtain \eqref{eq:FL-converse}.
\QED

\subsection{Proof of Theorems \ref{theo:d-fixed-length_resolvability} and \ref{theo:alt_fixed-length_resolvability}}

The relations shown in \eqref{eq:general_formula_relation2} and \eqref{eq:opt_general_formula_relation2} imply that to prove Theorems \ref{theo:d-fixed-length_resolvability} and \ref{theo:alt_fixed-length_resolvability}, it suffices to show 
\begin{align}
\FLS(\delta| \vect{X}, \vect{W}) &\le \inf_{\hat{\vect{X}} \in B_\delta (\vect{X}, \vect{W})} \overline{I} (\hat{\vect{X}}; \hat{\vect{Y}}), \label{eq:direct_part_show}\\
\OFLS(\delta| \vect{X}, \vect{W}) &\le \inf_{\hat{\vect{X}} \in B_\delta^* (\vect{X}, \vect{W})} \overline{I} (\hat{\vect{X}}; \hat{\vect{Y}}) \label{eq:opt_direct_part_show}
\end{align}
in the direct (achievability) part and 
\begin{align}
\FLS(\delta| \vect{X}, \vect{W}) &\ge \inf_{\hat{\vect{X}} \in B_\delta (\vect{X}, \vect{W})} \sup_{\substack{\varepsilon \ge 0, \\ \vect{Z} \in \tilde{B}_{\varepsilon} (\hat{\vect{Y}})}}   \overline{D}_{\varepsilon} (\vect{W}|| \vect{Z} | \hat{\vect{X}}), \label{eq:converse_part_show} \\
\OFLS(\delta| \vect{X}, \vect{W}) &\ge \inf_{\hat{\vect{X}} \in B_\delta^* (\vect{X}, \vect{W})} \sup_{\substack{\varepsilon \ge 0, \\ \vect{Z} \in \tilde{B}_{\varepsilon} (\hat{\vect{Y}})}}   \overline{D}_{\varepsilon} (\vect{W}|| \vect{Z} | \hat{\vect{X}}) \label{eq:opt_converse_part_show}
\end{align}
in the converse part.

1) \emph{Direct part:}~~
First, fix $\gamma > 0$ arbitrarily.
 Setting 
\begin{align}
R &= \inf_{\hat{\vect{X}} \in B_\delta (\vect{X}, \vect{W})} \overline{I} (\hat{\vect{X}}; \hat{\vect{Y}}) + 3 \gamma,  \label{eq:R_setting}
\end{align}
we show that $R$ is $\delta$-achievable, which means \eqref{eq:direct_part_show}.

Let $\vect{V} = \{V^n\}_{n=1}^\infty$ be a general source satisfying $\vect{V} \in  B_\delta (\vect{X}, \vect{W})$ and
\begin{align}
\overline{I} (\vect{V}; \vect{Z}) \le \inf_{\hat{\vect{X}} \in  B_\delta (\vect{X}, \vect{W})}  \overline{I} (\hat{\vect{X}}; \hat{\vect{Y}}) + \gamma, \label{eq:good_Z}
\end{align}
where $\vect{Z} = \{ Z^n\}_{n=1}^\infty$ denotes the output process via $\vect{W}$ due to the input process $\vect{V}$.
Setting $M_n = e^{n(  \overline{I} (\vect{V}; \vect{Z}) + 2 \gamma)}$, it follows from \eqref{eq:R_setting} and \eqref{eq:good_Z} that 
\begin{align}
\hspace*{-2mm}\limsup_{n \rightarrow \infty} \frac{1}{n} \log M_n &=   \overline{I} (\vect{V}; \vect{Z}) + 2 \gamma \le R.  \label{eq:rate_cond_satisfy1}
\end{align}

Lemma \ref{lem:achievability_bound} with $c =   \overline{I}(\vect{V}; \vect{Z}) + \gamma$ guarantees the existence of a deterministic mapping $\varphi_n: \{ 1, 2, \ldots, M_n \} \rightarrow \mathcal{X}^n$ with the uniform random number ${U_{M_n}}$ satisfying 
\begin{align}
  &\hspace*{-2mm} \limsup_{n \rightarrow \infty} d(P_{Z^n}, P_{\tilde{Y}^n}) \nonumber \\
  &\le \limsup_{n \rightarrow \infty} \Pr \left\{ \frac{1}{n} \log \! \frac{W^n(Z^n | V^n)}{P_{Z^n}(Z^n)} >    \overline{I}(\vect{V}; \vect{Z}) + \gamma \right\} \nonumber \\
   &= 0, \label{eq:variational_dist_cond1a}
\end{align} 
where $\tilde{Y}^n$ denotes the output via $W^n$ due to the input $\tilde{X}^n = \varphi_n({U_{M_n}})$.
Then, the triangle inequality leads to
\begin{align}
 &\hspace*{-2mm} \limsup_{n \rightarrow \infty}  d(P_{Y^n}, P_{\tilde{Y}^n}) \nonumber \\
&\hspace*{-2mm}~ \le \limsup_{n \rightarrow \infty} d(P_{Y^n}, P_{Z^n}) + \lim_{n \rightarrow \infty} d(P_{Z^n}, P_{\tilde{Y}^n}) \le \delta, \label{eq:variational_dist_cond_satisfy1}
\end{align}
where the last inequality is due to the fact $\vect{V} \in  B_\delta (\vect{X}, \vect{W})$ and \eqref{eq:variational_dist_cond1a}.
Combining \eqref{eq:rate_cond_satisfy1} and \eqref{eq:variational_dist_cond_satisfy1} concludes that $R$ is $\delta$-achievable, and hence \eqref{eq:direct_part_show} holds.

To prove \eqref{eq:opt_direct_part_show}, for any given $\gamma > 0$ setting 
\begin{align}
R &= \inf_{\hat{\vect{X}} \in B_\delta^* (\vect{X}, \vect{W})} \overline{I} (\hat{\vect{X}}; \hat{\vect{Y}}) + 3 \gamma, 
\end{align}
we show that $R$ is optimistically $\delta$-achievable. 
Let $\vect{V} = \{V^n\}_{n=1}^\infty$ be a general source satisfying $\vect{V} \in  B_\delta^* (\vect{X}, \vect{W})$ and
\begin{align}
\overline{I} (\vect{V}; \vect{Z}) \le \inf_{\hat{\vect{X}} \in  B_\delta^* (\vect{X}, \vect{W})}  \overline{I} (\hat{\vect{X}}; \hat{\vect{Y}}) + \gamma, \label{eq:opt_good_Z}
\end{align}
where $\vect{Z} = \{ Z^n\}_{n=1}^\infty$ denotes the output process via $\vect{W}$ due to input $\vect{V}$.
Along the same line to prove \eqref{eq:direct_part_show}, it is easily verified that there exists a deterministic mapping  $\varphi_n: \{ 1, 2, \ldots, M_n \} \rightarrow \mathcal{X}^n$ satisfying \eqref{eq:rate_cond_satisfy1} and \eqref{eq:variational_dist_cond1a}.
Then, the triangle inequality leads to
\begin{align}
 &\hspace*{-2mm} \liminf_{n \rightarrow \infty}  d(P_{Y^n}, P_{\tilde{Y}^n}) \nonumber \\
&\hspace*{-2mm}~ \le \liminf_{n \rightarrow \infty} d(P_{Y^n}, P_{Z^n}) + \lim_{n \rightarrow \infty} d(P_{Z^n}, P_{\tilde{Y}^n}) \le \delta, \label{eq:variational_dist_cond_satisfy2}
\end{align}
where the last inequality is due to the fact $\vect{V} \in  B_\delta^* (\vect{X}, \vect{W})$.
Combining \eqref{eq:rate_cond_satisfy1} and \eqref{eq:variational_dist_cond_satisfy2} concludes that $R$ is optimistically $\delta$-achievable, and hence \eqref{eq:opt_direct_part_show} holds.
\QED

\medskip
2) \emph{Converse part:}~~
We shall prove \eqref{eq:converse_part_show} and \eqref{eq:opt_converse_part_show} to establish the converse part of Theorems \ref{theo:d-fixed-length_resolvability} and \ref{theo:alt_fixed-length_resolvability}.

Let $R$ be $\delta$-achievable.
Then, there exists a mapping $\varphi_n : \{ 1, 2, \ldots, M_n\} \rightarrow \mathcal{X}^n$ satisfying \eqref{eq:rate_cond} and \eqref{eq:variational_dist_cond}.
Let $\gamma > 0$ be fixed arbitrarily.
From \eqref{eq:rate_cond}, we have
\begin{align}
\frac{1}{n} \log M_n \le R + \gamma \label{eq:ineq10b}
\end{align}
for all sufficiently large $n$.
Fixing an $\varepsilon \in [0,1)$ arbitrarily, we choose any $\vect{Z} \in \tilde{B}_\varepsilon (\tilde{\vect{Y}})$, where $\tilde{\vect{Y}} =\{ \tilde{Y}^n \}_{n=1}^\infty$ denotes the output via $\vect{W}$ due to input $\tilde{\vect{X}}=\{ \tilde{X}^n  = \varphi_n (U_{M_n})\}_{n=1}^\infty$.
By using Lemma \ref{lem:FL-converse} with $c = \frac{1}{n} \log M_n + \gamma$ and \eqref{eq:ineq10b}, we have
\begin{align}
 &\hspace*{-1mm} d(P_{Z^n}, P_{\tilde{Y}^n}) \nonumber \\
&\hspace*{-1mm}~\ge \Pr  \left\{ \frac{1}{n} \log \frac{W^n(\tilde{Y}^n | \tilde{X}^n)}{P_{Z^n}(\tilde{Y}^n)} > R + 2 \gamma \right\} - e^{-n\gamma} \label{eq:LB1} 
\end{align}
for all sufficiently large $n$. 
 Since $\vect{Z} \in \tilde{B}_\varepsilon (\tilde{\vect{Y}})$, we obtain
\begin{align}
\limsup_{n \rightarrow \infty} \Pr  \left\{ \frac{1}{n} \log \frac{W^n(\tilde{Y}^n | \tilde{X}^n)}{P_{Z^n}(\tilde{Y}^n)} > R + 2 \gamma \right\} \le \varepsilon. \label{eq:asymptotic_prob}
\end{align}
Since $\varepsilon \in [0,1)$ and $\vect{Z} \in \tilde{B}_\varepsilon (\tilde{\vect{Y}})$ have been fixed arbitrarily, \eqref{eq:asymptotic_prob} implies
\begin{align}
R + 2 \gamma &\ge  \sup_{\substack{\varepsilon \ge 0, \\ \vect{Z} \in \tilde{B}_\varepsilon (\tilde{\vect{Y}})}} \overline{D}_\varepsilon (\vect{W}||\vect{Z}|\tilde{\vect{X}}). 
\end{align}
Since $\gamma >0 $ is arbitrary and $\tilde{\vect{X}} \in B_\delta (\vect{X}, \vect{W})$ follows from \eqref{eq:opt_variational_dist_cond}, we obtain
\begin{align}
R \ge \inf_{\hat{\vect{X}} \in B_\delta (\vect{X}, \vect{W})}  \sup_{\substack{\varepsilon \ge 0, \\ \vect{Z} \in \tilde{B}_\varepsilon (\hat{\vect{Y}})}}  \overline{D}_\varepsilon (\vect{W}||\vect{Z}|\hat{\vect{X}}), \end{align}
where $\hat{\vect{Y}} =\{ \awidehat{Y}^n \}_{n=1}^\infty$ denotes the output via $\vect{W}$ due to input $\hat{\vect{X}}=\{ \awidehat{X}^n\}$.
Thus, we obtain  \eqref{eq:converse_part_show}.

The proof of \eqref{eq:opt_converse_part_show} is analogous by using the fact $\tilde{\vect{X}} \in B_\delta^* (\vect{X}, \vect{W})$, completing the proof of the converse parts.
\QED

\section{Source Resolvability: Revisited}

When the channel $W^n$ is an identity mapping, the addressed problem reduces to the problem of \emph{source resolvability} \cite{Han2003}, where the target distribution is the general source $X^n$ itself.
In this case, we denote $\FLS (\delta | \vect{X}, \vect{W})$ simply by $\FLS (\delta | \vect{X})$.
For this problem, Steinberg and Verd\'u \cite{Steinberg-Verdu1996} have shown the following theorem, which generalizes the resolvability theorem established by Han and Verd\'u \cite{Han-Verdu1993} for $\delta = 0$:
\begin{e_theo}[Han and Verd\'u \cite{Han-Verdu1993}, Steinberg and Verd\'u \cite{Steinberg-Verdu1996}] \label{theo:source_resolvability}
For any target general source $\vect{X}$, 
\begin{align}
\FLS(\delta | \vect{X}) = \overline{H}_\delta (\vect{X}) ~~~(\delta \in [0,1)),
\end{align}
where
\begin{align}
\overline{H}_\delta (\vect{X}) &:= \mbox{\rm $\delta$p-}\limsup_{n \rightarrow \infty}  \frac{1}{n} \log \frac{1}{P_{X^n}(X^n)}
\end{align}
is the $\delta$-spectral sup-entropy rate for $\vect{X}$.
\QED
\end{e_theo}

When the channel $W^n$ is an identity mapping, we have $\overline{I} (\vect{X}; \vect{Y}) = \overline{H} (\vect{X})$
because
\begin{align}
\frac{1}{n} \log \frac{W^n(Y^n|X^n)}{P_{Y^n}(Y^n)} = \frac{1}{n} \log \frac{1}{P_{X^n}(X^n)}~~~\mbox{a.s.}
\end{align}
The following relation can be obtained from Theorems \ref{theo:d-fixed-length_resolvability} and \ref{theo:source_resolvability}, which gives a new characterization for $\overline{H}_\delta (\vect{X})$ and 
\begin{align}
\overline{H}_\delta^* (\vect{X}) &:= \mbox{\rm $\delta$p$^*$-}\limsup_{n \rightarrow \infty}  \frac{1}{n} \log \frac{1}{P_{X^n}(X^n)}.
\end{align}
\begin{e_theo} \label{theo:sup_entropy_relation}
For any general source $\vect{X}$,
\begin{align}
\overline{H}_\delta (\vect{X}) &= \inf_{\hat{\vect{X}} \in \tilde{B}_\delta (\vect{X})} \overline{H}(\hat{\vect{X}}), \label{eq:sup_entropy_relation} \\
\overline{H}_\delta^* (\vect{X}) &= \inf_{\hat{\vect{X}} \in \tilde{B}_\delta^* (\vect{X})} \overline{H}(\hat{\vect{X}})\label{eq:sup_entropy_relation2} 
\end{align}
for all $\delta \in [0,1)$, where
\begin{align}
\tilde{B}_\delta (\vect{X}) &:= \left\{ \hat{\vect{X}} =\big\{ \awidehat{X}^n \big\}_{n=1}^\infty : \,  \limsup_{n \rightarrow \infty} d(P_{X^n}, P_{\awidehat{X}^n}) \le \delta \right\}, \nonumber \\
\tilde{B}_\delta^* (\vect{X}) &:= \left\{ \hat{\vect{X}} =\big\{ \awidehat{X}^n \big\}_{n=1}^\infty : \,  \liminf_{n \rightarrow \infty} d(P_{X^n}, P_{\awidehat{X}^n}) \le \delta \right\}. \nonumber 
\end{align}
\QED
\end{e_theo}
Equations \eqref{eq:sup_entropy_relation} and \eqref{eq:sup_entropy_relation2} indicate that $\overline{H}_\delta (\vect{X}) $ and $\overline{H}_\delta^* (\vect{X}) $  can be viewed as ``smoothed'' 0-spectral sup-entropy rates.
These equations can also be proven directly from the property of the $\delta$-spectral sup-entropy rates $\overline{H}_\delta (\vect{X})$ and $\overline{H}_\delta^* (\vect{X})$, respectively.

\section{Application of General Formulas to Memoryless Source and Channel}

Now, let us consider a special case, where $\mathcal{X}$ and $\mathcal{Y}$ are finite sets and for each $n = 1, 2, \cdots$, both $X^n$ and $W^n$ are memoryless with joint probability
\begin{align}
P_{X^n}(\vect{x})W^n(\vect{y} | \vect{x}) = \left\{ \!\!
\begin{array}{ll}
\prod_{i=1}^n P_{X_1}(x_i)W_1(y_i|x_i) & \! \mbox{\rm for~odd~$n$}\\
\prod_{i=1}^n P_{X_2}(x_i)W_2(y_i|x_i)  & \! \mbox{\rm for~even~$n$} \\
\end{array}
\right.  \nonumber
\end{align}
for $\vect{x}=(x_1, \ldots, x_n) \in \mathcal{X}^n$ and $\vect{y}=(y_1, \ldots, y_n) \in \mathcal{Y}^n$, where $X_j$ and $W_j$ $(j=1, 2)$ denote a source and a channel, respectively.
The source $\vect{X}=\{X^n\}_{n=1}^\infty$ and the channel $\vect{W}=\{W^n\}_{n=1}^\infty$ are completely characterized by $P_{X_1}W_1$ if $n$ is odd and by $P_{X_2}W_2$ if $n$ is even and are known as one of the simplest examples for which $\FLS(\delta|X, W)$ and $\OFLS(\delta|X, W)$ do not coincide in general \cite{Han2003}.
Let $Y_j$ denote the output via $W_j$ due to input $X_j$ for $j=1,2$.
The alternative formulas \eqref{eq:d-fixed-length_general_formula2} and \eqref{eq:opt_d-length_general_formula2} are of use to prove the converse parts.

\begin{e_theo}\label{theo:opt_DMC-fixed-length_resolvability}
For any $\delta \in [0,1)$,
\begin{align}
\hspace*{-3mm} \FLS (\delta | X, W) 
&=\max_{j=1,2} \inf_{\awidehat{X}_j \in B_0 (X_j, W_j)} I (\awidehat{X}_j; \awidehat{Y}_j), \label{eq:DMC-fixed-length_formula1} \\
\hspace*{-3mm}\OFLS (\delta | X, W) 
&= \min_{j=1,2} \inf_{\awidehat{X}_j \in B_0 (X_j, W_j)} I (\awidehat{X}_j; \awidehat{Y}_j), \label{eq:DMC-fixed-length_formula2}
\end{align}
where $\awidehat{Y}_j$ denotes the output via $W_j$ due to the input $\awidehat{X}_j$, $I(\awidehat{X}_j; \awidehat{Y}_j)$  denotes the mutual information between $\awidehat{X}_j$ and $\awidehat{Y}_j$,  and we define $B_0 (X_j, W_j) := \big\{ \awidehat{X}_j: P_{Y_j} = P_{\awidehat{Y}_j} \big\}$.
\end{e_theo}
(\emph{Proof}) The proof is given in \ref{append:proof_theorem_perturb}.
\QED

It should be noticed that the constant $\delta$ does not appear in formulas \eqref{eq:DMC-fixed-length_formula1} and \eqref{eq:DMC-fixed-length_formula2}. This result indicates that the \emph{strong converse} holds for the memoryless source and channel pair. 
Precisely, for any
\begin{align}
R < \min_{j=1,2} \inf_{\awidehat{X}_j \in B_0 (X_j, W_j)} I (\awidehat{X}_j; \awidehat{Y}_j), 
\end{align}
any mapping $\varphi_n: \{ 1, \ldots, M_n\} \rightarrow \mathcal{X}^n$ satisfying \eqref{eq:opt_rate_cond} produces the variational distance $d(P_{Y^n}, P_{\tilde{Y}^n}) \rightarrow 1~(n \rightarrow \infty)$, where $\tilde{Y}^n$ denotes the output via $W^n$ due to input $\tilde{X}^n = \varphi_n(U_{M_n})$.

For an i.i.d.\ source  $X$ with $X=X_1=X_2$  and a stationary memoryless channel $W$ with $W=W_1=W_2$, we obtain the following corollary from Theorem \ref{theo:opt_DMC-fixed-length_resolvability}, which has been proved by Watanabe and Hayashi \cite{Watanabe-Hayashi2014}.
\begin{e_coro}[Watanabe and Hayashi \cite{Watanabe-Hayashi2014}] \label{coro:DMC-fixed-length_resolvability}
{\rm
For any i.i.d.\ input source $X$ and any stationary memoryless channel $W$, 
\begin{align}
\hspace*{-1mm} \FLS(\delta | X, W) =\OFLS (\delta | X, W) 
&=\inf_{\awidehat{X} \in B_0 (X, W)} I (\awidehat{X}; \awidehat{Y}) \label{eq:DMC-fixed-length_formula}
\end{align}
for every $\delta \in [0,1)$, 
where $\awidehat{Y}$ denotes the output via $W$ induced by input $\awidehat{X}$.
}\QED
\end{e_coro}

\section{Second-Order Channel Resolvability}


We turn to considering the \emph{second-order} resolution rates \cite{Watanabe-Hayashi2014}. First, we define the second-order achievability.

\begin{e_defin}[$(\delta, R)$-Channel Resolvability]
{\rm
Let $\delta \in [0,1)$ and $R \ge 0$ be fixed arbitrarily.
A resolvability rate $L $ is said to be \emph{$(\delta, R)$-achievable at $\vect{X}$} if there exists a deterministic mapping $\varphi_n : \{ 1, \ldots, M_n\} \rightarrow \mathcal{X}^n$ satisfying
\begin{align}
 \limsup_{n \rightarrow \infty}  \frac{1}{\sqrt{n}} \left( \log M_n -nR \right) &\le L, \label{eq:2nd_order_rate_cond}  \\ 
 \limsup_{n \rightarrow \infty} d(P_{Y^n}, P_{\tilde{Y}^n}) &\le \delta \label{eq:2nd_order_variational_dist_cond},
\end{align}
where $\tilde{Y}^n$ denotes the output via $W^n$ due to the input $\tilde{X}^n = \varphi_n({U_{M_n}})$.
We define
\begin{align}
T (\delta, R | \vect{X}, \vect{W}):= \inf \{ L : ~ L ~\mbox{is $(\delta, R)$-achievable at}~\vect{X} \}, \nonumber 
\end{align}
which is called the \emph{$(\delta, R)$-channel resolvability} (at $\vect{X})$.}
\QED
\end{e_defin}

As in the first-order case, we address the relaxed constraint on the variational distance.
\begin{e_defin}[Optimistic $(\delta, R)$-Channel Resolvability]
{\rm
Let $\delta \in [0,1)$ and $R \ge 0$ be fixed arbitrarily.
A resolvability rate $L $ is said to be \emph{optimistically $(\delta, R)$-achievable at $\vect{X}$} if there exists a deterministic mapping $\varphi_n : \{ 1, \ldots, M_n\} \rightarrow \mathcal{X}^n$ satisfying
\begin{align}
 \limsup_{n \rightarrow \infty}  \frac{1}{\sqrt{n}} \left( \log M_n -nR \right) &\le L, \label{eq:opt_2nd_order_rate_cond}  \\ 
 \liminf_{n \rightarrow \infty} d(P_{Y^n}, P_{\tilde{Y}^n}) &\le \delta \label{eq:opt_2nd_order_variational_dist_cond},
\end{align}
where $\tilde{Y}^n$ denotes the output via $W^n$ due to the input $\tilde{X}^n = \varphi_n({U_{M_n}})$.
We define
\begin{align}
&T^* (\delta, R | \vect{X}, \vect{W}) \nonumber \\
&~~:= \inf \{ L : ~ L ~\mbox{is optimistically $(\delta, R)$-achievable at}~\vect{X} \}, \nonumber 
\end{align}
called the \emph{optimistic $(\delta, R)$-channel resolvability} (at $\vect{X})$.}
\QED
\end{e_defin}

\begin{e_rema} \label{rema:interesting_second_order}
By definition, it is easily verified that 
\begin{align}
T (\delta, R|{\vect{X}},\vect{W})  = \left\{
\begin{array}{ll}
+  \infty & \mbox{for}~ R < \FLS (\delta|{\vect{X}}, \vect{W})  \\
-  \infty & \mbox{for}~ R > \FLS(\delta|{\vect{X}}, \vect{W}).
\end{array}
\right.
\end{align}
Hence, only the case $R = \FLS(\delta|{\vect{X}}, \vect{W})$ is of our interest.
Similarly, when discussing the optimistic $(\delta,R)$-channel resolvability, the case $R = \OFLS(\delta|{\vect{X}}, \vect{W})$ is our primary interest.

\QED
\end{e_rema}

Now, we establish the general formulas for the second-order resolvability.
The following two theorems can be proven analogously to Theorems \ref{theo:d-fixed-length_resolvability} and \ref{theo:alt_fixed-length_resolvability} in the first-order case.

\begin{e_theo}\label{theo:2nd_oder_fixed-length_resolvability}
Let  $\delta \in [0,1)$ and $R \ge 0$ be fixed arbitrarily.
For any input process $\vect{X}$ and any general channel $\vect{W}$, 
\begin{align}
T (\delta, R | \vect{X}, \vect{W}) 
&=\inf_{\hat{\vect{X}} \in B_\delta (\vect{X}, \vect{W})} \overline{I} (R|\hat{\vect{X}}; \hat{\vect{Y}}), \label{eq:2nd-fixed-length_general_formula1} \\
T^* (\delta, R | \vect{X}, \vect{W}) 
&=\inf_{\hat{\vect{X}} \in B_\delta^* (\vect{X}, \vect{W})} \overline{I} (R|\hat{\vect{X}}; \hat{\vect{Y}}), \label{eq:opt_2nd-length_general_formula1} 
\end{align}
where $\hat{\vect{Y}} = \{ \awidehat{Y}^n\}_{n=1}^\infty$ denotes the output process via $\vect{W}$ due to the input process $\hat{\vect{X}} = \{ \awidehat{X}^n\}_{n=1}^\infty$, and we define
\begin{align}
\overline{I} (R|\vect{X}; \vect{Y}) :=   \mbox{\rm p-}\limsup_{n \rightarrow \infty}  \frac{1}{\sqrt{n}} \left( \log \frac{W^n(Y^n|X^n)}{P_{Y^n}(Y^n)} - nR \right). \nonumber
\end{align}
\QED
\end{e_theo}

We give alternative formulas in the following theorem, which correspond to Theorem \ref{theo:alt_fixed-length_resolvability} on the first-order resolvability rates:
\begin{e_theo}\label{theo:2nd_alt_fixed-length_resolvability}
Let  $\delta \in [0,1)$ and $R > 0$ be fixed arbitrarily.
For any input process $\vect{X}$ and any general channel $\vect{W}$, 
\begin{align}
T (\delta, R | \vect{X}, \vect{W}) &=\inf_{\hat{\vect{X}} \in B_\delta (\vect{X}, \vect{W})} \sup_{\substack{\varepsilon \ge 0, \\ \vect{Z} \in \tilde{B}_{\varepsilon} (\hat{\vect{Y}})}}   
\overline{J}_{\varepsilon} (R|\vect{W}, \vect{Z}, \hat{\vect{X}}), \label{eq:2nd-fixed-length_general_formula2} \\
T^* (\delta, R | \vect{X}, \vect{W}) &=\inf_{\hat{\vect{X}} \in B_\delta^* (\vect{X}, \vect{W})} \sup_{\substack{\varepsilon \ge 0, \\ \vect{Z} \in \tilde{B}_{\varepsilon} (\hat{\vect{Y}})}} 
\overline{J}_{\varepsilon} (R| \vect{W}, \vect{Z} , \hat{\vect{X}}), \label{eq:opt_2nd-length_general_formula2}
\end{align}
where $\hat{\vect{Y}} = \{ \awidehat{Y}^n\}_{n=1}^\infty$ denotes the output process via $\vect{W}$ due to input process $\hat{\vect{X}} = \{ \awidehat{X}^n\}_{n=1}^\infty$, and we define
\begin{align}
&\overline{J}_\varepsilon (R| \vect{W}, \vect{Z}, \hat{\vect{X}}) \nonumber \\
&~~ :=  \mbox{\rm $\varepsilon$p-}\limsup_{n \rightarrow \infty} \frac{1}{\sqrt{n}} \left( \log \frac{W^n(\awidehat{Y}^n|\awidehat{X}^n)}{P_{Z^n}(\awidehat{Y}^n)} - nR \right). \nonumber 
\end{align}
\QED
\end{e_theo}

When the channel is an identity mapping, the problem addressed here reduces to finding the second-order $\delta$-\emph{source} resolvability \cite{Nomura-Han2013}.
In this case, we denote $T (\delta, R | \vect{X}, \vect{W}) $ simply by $T (\delta, R | \vect{X}) $.
Nomura and Han \cite{Nomura-Han2013} have established the following fundamental theorem, which generalizes the theorem on the  first-order $\delta$-source resolvability given by \cite{Han-Verdu1993,Steinberg-Verdu1996}:
\begin{e_theo}[Nomura and Han \cite{Nomura-Han2013}] \label{theo:2nd_source_resolvability}
For any target general source $\vect{X}$, 
\begin{align}
\FLS(\delta, R | \vect{X}) = \overline{H}_\delta (R|\vect{X}) ~~~(\delta \in [0,1)),
\end{align}
where 
\begin{align}
\overline{H}_\delta (R|\vect{X}) &:= \mbox{\rm $\delta$p-}\limsup_{n \rightarrow \infty}  \frac{1}{\sqrt{n}} \left(\log \frac{1}{P_{X^n}(X^n)} -nR \right). \nonumber
\end{align}
\QED
\end{e_theo}

Since the channel $W^n$ is the identity mapping, we have $\overline{I} (R|\vect{X}; \vect{Y}) = \overline{H} (R|\vect{X})$.
The following relation can be obtained from Theorems \ref{theo:2nd_oder_fixed-length_resolvability} and \ref{theo:2nd_source_resolvability}, which gives a new representation for $\overline{H}_\delta (R|\vect{X})$ and 
\begin{align}
\overline{H}_\delta^* (R| \vect{X}) &:= \mbox{\rm $\delta$p$^*$-}\limsup_{n \rightarrow \infty} \frac{1}{\sqrt{n}} \left(\log \frac{1}{P_{X^n}(X^n)} -nR \right). \nonumber
\end{align}
\begin{e_theo} \label{theo:2nd_sup_entropy_relation}
For any general source $\vect{X}$,
\begin{align}
\overline{H}_\delta (R|\vect{X}) &= \inf_{\hat{\vect{X}} \in \tilde{B}_\delta (\vect{X})} \overline{H}(R|\hat{\vect{X}}), \label{eq:2nd_sup_entropy_relation} \\
\overline{H}_\delta^* (R|\vect{X}) &= \inf_{\hat{\vect{X}} \in \tilde{B}_\delta^* (\vect{X})} \overline{H}(R|\hat{\vect{X}})\label{eq:2nd_sup_entropy_relation2} 
\end{align}
for all $\delta \in [0,1)$ and $R \ge 0$, where we define $\overline{H}(R|\hat{\vect{X}}) = \overline{H}_0(R|\hat{\vect{X}})$ and $\overline{H}^*(R|\hat{\vect{X}}) = \overline{H}_0^* (R|\hat{\vect{X}})$. 
\QED
\end{e_theo}
Equation \eqref{eq:2nd_sup_entropy_relation2}  as well as \eqref{eq:2nd_sup_entropy_relation} can be proven directly from the definition of the quantities on both sides.
As was shown in \eqref{eq:sup_entropy_relation} and \eqref{eq:sup_entropy_relation2} in the first-order case, so-called smoothing operations appear here; both $\overline{H}_\delta (R|\vect{X}) $ and $\overline{H}_\delta^* (R|\vect{X}) $  are characterized by $\overline{H} (R|\hat{\vect{X}})$ of  a general source $\hat{\vect{X}}$ in the $\delta$-ball $ \tilde{B}_\delta (\vect{X})$ and $ \tilde{B}_\delta^* (\vect{X})$ centered at $\vect{X}$, respectively.



%
%


\appendices

\section{Proof of Theorem \ref{theo:opt_DMC-fixed-length_resolvability}} \label{append:proof_theorem_perturb}

\noindent
1) \emph{Direct part:}

Without loss of generality, we assume that
\begin{align}
 \inf_{\awidehat{X}_1 \in B (X_1, W_1)} I (\awidehat{X}_1; \awidehat{Y}_1)  \ge  \inf_{\awidehat{X}_2 \in B (X_2, W_2)} I (\awidehat{X}_2; \awidehat{Y}_2).  \label{eq:ordered_system}
\end{align}

\noindent
(i)~~
First, fix $\gamma > 0$ arbitrarily. For $j=1,2$, let $\overline{X}_j^n$ be $n$ i.i.d.\ samples from source $P_{\overline{X}_j}$ satisfying $\overline{X}_j \in  B (X_j, W_j)$ and
\begin{align}
I (\overline{X}_j; \overline{Y}_j) \le \inf_{\awidehat{X}_j \in  B (X_j, W_j)}  I (\awidehat{X}_j; \awidehat{Y}_j) + \gamma, \label{eq:MI_UB}
\end{align}
where $\overline{Y}_j$ denotes the output via $W_j$ due to input $\overline{X}_j$.
Set $(\Vn, \Zn) = (\overline{X}_1^n, \overline{Y}_1^n)$ for odd $n$
and $(\Vn, \Zn) = (\overline{X}_2^n, \overline{Y}_2^n)$ for even $n$.
Since the random variable 
\begin{align}
\frac{1}{n} \log \frac{W^n(\Zn | \Vn)}{P_{\Zn}(\Zn)} = \frac{1}{n} \sum_{i=1}^k \log \frac{W(Z_i | V_i)}{P_{Z_i}(Z_i)}
\end{align}
 is a sum of independent random variables, where $\Vn = (V_1, V_2, \ldots, V_n)$ and  $\Zn = (Z_1, Z_2, \ldots, Z_n)$, its expected value satisfies
\begin{align}
\E \left\{\frac{1}{n} \log \frac{W^n(\Zn | \Vn)}{P_{\Zn}(\Zn)}  \right\} \le I(\overline{X}_1; \overline{Y}_1) + \gamma ~~~\mathrm{for~all~} n. \nonumber 
\end{align} 
The weak law of large numbers guarantees
\begin{align}
\lim_{n \rightarrow \infty} \Pr \left\{ \frac{1}{n} \log \frac{W^n(\Zn | \Vn)}{P_{\Zn}(\Zn)} >  I(\overline{X}_1; \overline{Y}_1) + \gamma \right\}  = 0,\label{eq:spectrum_cond1}
\end{align}
which indicates that
\begin{align}
\overline{I} (\vectV; \vectZ) \le I(\overline{X}_1; \overline{Y}_1) + \gamma, \label{eq:spectrum_UB1}
\end{align}
where $\vectV = \{ \Vn\}_{n=1}^\infty$ and $\vectZ = \{ \Zn\}_{n=1}^\infty$.
On the other hand, because it obviously holds that $\vectV \in B_\delta (\vect{X}, \vect{W})$, we have
\begin{align}
\inf_{\hat{\vect{X}} \in B_\delta (\vect{X}, \vect{W})}  \overline{I} (\hat{\vect{X}}; \hat{\vect{Y}}) \le \overline{I} (\vectV; \vectZ). \label{eq:spectrum_UB2}
\end{align}
Since $\gamma > 0$ is an arbitrary constant,  \eqref{eq:MI_UB}, \eqref{eq:spectrum_UB1} and \eqref{eq:spectrum_UB2} imply
\begin{align}
\inf_{\hat{\vect{X}} \in B_\delta (\vect{X}, \vect{W})}  \overline{I} (\hat{\vect{X}}; \hat{\vect{Y}}) \le \inf_{\awidehat{X}_1 \in B (X_1, W_1)} I(\hat{X}_1; \hat{Y}_1) .  \label{eq:spectrum_UB3}
\end{align}

\noindent
(ii)~~For an arbitrary fixed $\gamma > 0$, let $\overline{X}_2^n$ be $n$ i.i.d.\ samples from source $P_{\overline{X}_2}$ satisfying $\overline{X}_2 \in  B (X_2, W_2)$ and
\eqref{eq:MI_UB} with $j = 2$.
Also, let $\overline{X}_1^n$ be $n$ i.i.d.\ samples from source $P_{\overline{X}_1}$ satisfying $I(\overline{X}_1; \overline{Y}_1) = 0$, where $\overline{Y}_1$ denotes the output via $W_1$ due to input $\overline{X}_1$.
Set $(\Vn, \Zn) = (\overline{X}_1^n, \overline{Y}_1^n)$ for odd $n$
and $(\Vn, \Zn) = (\overline{X}_2^n, \overline{Y}_2^n)$ for even $n$.
Then, we obtain
\begin{align}
\E \left\{\frac{1}{n} \log \frac{W^n(\Zn | \Vn)}{P_{\Zn}(\Zn)}  \right\} \le I(\overline{X}_2; \overline{Y}_2)  ~~~\mathrm{for~all~} n. \nonumber 
\end{align} 
Again, by the weak law of large numbers, we have
\begin{align}
\lim_{n \rightarrow \infty} \Pr \left\{ \frac{1}{n} \log \frac{W^n(\Zn | \Vn)}{P_{\Zn}(\Zn)} >  I(\overline{X}_2; \overline{Y}_2) + \gamma \right\}  = 0,\label{eq:spectrum_cond2}
\end{align}
indicating that
\begin{align}
\overline{I} (\vectV; \vectZ) \le I(\overline{X}_2; \overline{Y}_2) + \gamma. \label{eq:spectrum_UB4}
\end{align}
On the other hand, it holds that $\vectV \in B_\delta^* (\vect{X}, \vect{W})$
because
\begin{align}
\liminf_{n \rightarrow \infty} d(P_{Y^n}, P_{\Zn}) \le \liminf_{k \rightarrow \infty} d(P_{Y^{2k}}, P_{Z^{2k}}) = 0.
\end{align}
Then, we have
\begin{align}
\inf_{\hat{\vect{X}} \in B_\delta^* (\vect{X}, \vect{W})}  \overline{I} (\hat{\vect{X}}; \hat{\vect{Y}}) \le \overline{I} (\vectV; \vectZ). \label{eq:spectrum_UB5}
\end{align}
Since $\gamma > 0$ is an arbitrary constant,  \eqref{eq:MI_UB} with $j=2$, \eqref{eq:spectrum_UB4} and \eqref{eq:spectrum_UB5} imply
\begin{align}
\inf_{\hat{\vect{X}} \in B_\delta^* (\vect{X}, \vect{W})}  \overline{I} (\hat{\vect{X}}; \hat{\vect{Y}}) \le \inf_{\awidehat{X}_2 \in B (X_2, W_2)} I(\hat{X}_2; \hat{Y}_2) .  \label{eq:spectrum_UB6}
\end{align}

\medskip
\noindent
2) \emph{Converse part:}

As was argued in \cite{Watanabe-Hayashi2014}, we shall use the method of types \cite{Csiszar-Korner2011}.  
The following notation is introduced.
\begin{itemize}

\item Let $P_{\vect{x}} $ denote the \emph{type} of $\vect{x}\in \mathcal{X}^n$, i.e., $P_{\vect{x}}(a)$ denotes the number of occurrence of symbol  $a \in \mathcal{X}$ in $\vect{x}$.

\item Let $P_{\vect{x}\vect{y}} $ denote the \emph{joint type} of $(\vect{x}, \vect{y}) \in \mathcal{X}^n \times \mathcal{Y}^n$.

\item Let $P_{\vect{x}}W(b) : = \sum_{a} P_{\vect{x}}(a) W(b|a)$ denote the marginal distribution on $\mathcal{Y}$.

\item Define the sets of \emph{$\varepsilon$-typical sequences} as
{\small
\begin{align}
T_{Y, \varepsilon}^n &:= \left\{ \vect{y} \in \mathcal{Y}^n : |P_{\vect{y}} (b) - P_{Y}(b)| \le  \varepsilon, ~\forall b \in \mathcal{Y} \right\},  \\
T_{W, \varepsilon}^n(\vect{x}) &:= \left\{ \vect{y} \in \mathcal{Y}^n : |P_{\vect{x}\vect{y}} (a,b) - P_{\vect{x}}(a)W(b|a)| \le  \varepsilon, \right. \nonumber \\
 & ~~~~~~~~~~~~~~~~~~\left. ~\forall (a,b) \in \mathcal{X} \times  \mathcal{Y} \right\},  \\
A_{Y}(\varepsilon) &:= \left\{ P \in \mathcal{P}(\mathcal{X}) : |PW (b) - P_{Y}(b)| \le  2|\mathcal{X}| \varepsilon,  \right. \nonumber \\
 & ~~~~~~~~~~~~~~~~~~~~~\left. ~\forall b \in \mathcal{Y} \right\}.  \label{eq:set_AY}
\end{align}
}
\end{itemize}

Now, we are in a position to prove the converse part of Theorem \ref{theo:opt_DMC-fixed-length_resolvability}.
We again assume \eqref{eq:ordered_system} without loss of generality.
In view of Theorems \ref{theo:d-fixed-length_resolvability} and \ref{theo:alt_fixed-length_resolvability}, we shall show
\begin{align}
&\inf_{\hat{\vect{X}} \in B_\delta (\vect{X}, \vect{W})} \sup_{\substack{\varepsilon \ge 0, \\ \vect{Z} \in \tilde{B}_{\varepsilon} (\hat{\vect{Y}})}}   \overline{D}_{\varepsilon} (\vect{W}|| \vect{Z} | \hat{\vect{X}}) \nonumber \\
&~~~~\ge \inf_{\hat{X}_1 \in B (X_1, W_1)} I (\hat{X}_1; \hat{Y}_1), \label{eq:converse_ineq1} \end{align}
and
\begin{align}
&\inf_{\hat{\vect{X}} \in B_\delta^* (\vect{X}, \vect{W})} \sup_{\substack{\varepsilon \ge 0, \\ \vect{Z} \in \tilde{B}_{\varepsilon} (\hat{\vect{Y}})}}   \overline{D}_{\varepsilon} (\vect{W}|| \vect{Z} | \hat{\vect{X}}) \nonumber \\
&~~~~\ge \inf_{\hat{X}_2 \in B (X_2, W_2)} I (\hat{X}_2; \hat{Y}_2).  \label{eq:converse_ineq2}
\end{align}

(i) To show \eqref{eq:converse_ineq1}, we first fix an arbitrary
\begin{align}
R > \inf_{\hat{\vect{X}} \in B_\delta (\vect{X}, \vect{W})} \sup_{\substack{\varepsilon \ge 0, \\ \vect{Z} \in \tilde{B}_{\varepsilon} (\hat{\vect{Y}})}}   \overline{D}_{\varepsilon} (\vect{W}|| \vect{Z} | \hat{\vect{X}}), \label{eq:rate_setting1}
\end{align}
and we shall show that $R$ is not smaller than the right-hand side of \eqref{eq:converse_ineq1}.
For simplicity, we define 
\begin{align}
j(n) = \left\{
\begin{array}{ll}
 1 & \mathrm{if~} n~\mathrm{ is~ odd} \\
 2 & \mathrm{if~} n~\mathrm{ is~ even}.
\end{array}
\right.  \label{eq:channel_index}
\end{align}
Then, we can write $X^n = X_{j(n)}^n$ and $W^n = W_{j(n)}^n$ and the corresponding output is $Y^n = Y_{j(n)}^n$.
Letting $\gamma > 0$ be arbitrarily fixed, we define $\gamma' := |\mathcal{X}| \gamma$, $\tau_n := \Pr \{ Y^n \in T_{Y_{j(n)}, \gamma'}^n \}$ and set the following probability distribution on $\mathcal{Y}^n$:
\begin{align}
P_{\overline{Y}^n} (\vect{y}) := \frac{P_{Y^n}(\vect{y}) \vect{1}\{ \vect{y} \in T_{Y_{j(n)}, \gamma'}^n \} }{\tau_n}~~~(\vect{y} \in \mathcal{Y}^n), \label{eq:overlineY}
\end{align}
where $\vect{1}\{E\}$ is the indicator function for the event $E$.
Then, from the property of the set of $\gamma'$-typical sequences $T_{Y_{j(n)}, \gamma'}^n$, we have $\tau_n \rightarrow 1$ as $n \rightarrow \infty$ and hence
\begin{align}
\lim_{n \rightarrow \infty} d(P_{Y^n}, P_{\overline{Y}^n}) =0. \label{eq:dist_convergence1}
\end{align}

Now, we can see that by \eqref{eq:rate_setting1} there exists an $\hat{\vect{X}} \in B_\delta (\vect{X}, \vect{W})$ satisfying
\begin{align}
R &> \sup_{\substack{\varepsilon \ge 0, \\ \vect{Z} \in \tilde{B}_{\varepsilon} (\hat{\vect{Y}})}}   \overline{D}_{\varepsilon} (\vect{W}|| \vect{Z} | \hat{\vect{X}}) - \gamma \nonumber \\
 & \ge \overline{D}_{\delta} (\vect{W}|| \overline{\vect{Y}} | \hat{\vect{X}}) - \gamma, \label{eq:LB2}
\end{align}
where $\hat{\vect{Y}} =\{\hat{Y}^n\}_{n=1}^\infty$ denotes the output via $\vect{W}$ due to input $\hat{\vect{X}} =\{\hat{X}^n\}_{n=1}^\infty$, and to derive \eqref{eq:LB2} we have used that fact that $\overline{\vect{Y}} \in \tilde{B}_{\delta} (\hat{\vect{Y}})$ which follows from $\hat{\vect{X}} \in B_\delta (\vect{X}, \vect{W})$ and \eqref{eq:dist_convergence1} with the triangle inequality:
\begin{align}
&\limsup_{n \rightarrow \infty} d(P_{\hat{Y}^n}, P_{\overline{Y}^n}) \nonumber \\
&~~~\le \lim_{n \rightarrow \infty} d(P_{Y^n}, P_{\overline{Y}^n}) + \limsup_{n \rightarrow \infty} d(P_{Y^n}, P_{\hat{Y}^n}) \le \delta.
\end{align}
We invoke the method of squeezing a subsequence of good types in the information spectrum approach as in \cite{YHN2016}. 
Equation \eqref{eq:LB2} implies that there exists some $\{ d_n > 0 : d_1 > d_2 > \cdots \rightarrow  \delta \}$ satisfying
\begin{align}
d_n &\ge \Pr  \left\{ \frac{1}{n} \log \frac{W^n(\hat{Y}^n | \hat{X}^n)}{P_{\overline{Y}^n}(\hat{Y}^n)} > R + 2 \gamma \right\}  \label{eq:ineq11}
\end{align}
for all $n = 1, 2, \cdots$.
Since 
\begin{align}
 & \Pr  \left\{ \frac{1}{n} \log \frac{W^n(\hat{Y}^n | \hat{X}^n)}{P_{\overline{Y}^n}(\hat{Y}^n)} > R + 2 \gamma \right\} \nonumber \\
 &~~~= \sum_{\vect{x} \in \mathcal{X}^n} P_{\hat{X}^n}(\vect{x}) W^n_{\vect{x}}  \left\{ \frac{1}{n} \log \frac{W^n(\hat{Y}^n | \vect{x})}{P_{\overline{Y}^n}(\hat{Y}^n)} > R + 2 \gamma  \right\} , \nonumber
\end{align}
where we use $W^n_{\vect{x}}$ to denote $W^n(\cdot | \vect{x})$ for simplicity,  \eqref{eq:ineq11} indicates that there exists some $\vect{x}_n \in \mathcal{X}^n$ satisfying 
\begin{align}
d_n &\ge  W^n_{\vect{x}_n}   \left\{ \frac{1}{n} \log \frac{W^n(\hat{Y}^n | \vect{x}_n)}{P_{\overline{Y}^n}(\hat{Y}^n)} > R + 2 \gamma \right\}  . \label{eq:ineq13}
\end{align} 
It is important to use the fact following from \eqref{eq:ineq13} that there exists a sequence of \emph{odd numbers} $\{ n_1 < n_2 < \cdots \rightarrow \infty \}$ and $P_\gamma \in \mathcal{P}(\mathcal{X})$ such that
\begin{align}
\limsup_{i \rightarrow \infty} d_{n_i} \le \delta,~~~~~
\lim_{i \rightarrow \infty} P_{\vect{x}_{n_i}} = P_\gamma, \label{eq:ineq14}
\end{align}
where $P_{\vect{x}_n}$ denotes the type of $\vect{x}_n \in \mathcal{X}^n$ (cf.\ \cite{YHN2016}).
The existence of such a convergent point $P_\gamma \in \mathcal{P}(\mathcal{X})$ follows from the fact that $ \mathcal{P}(\mathcal{X})$ is a compact set for finite $\mathcal{X}$. 
For notational simplicity, we use $k$ to denote (odd number) $k = n_1, n_2, \cdots$ so that \eqref{eq:ineq13} and \eqref{eq:ineq14} can be rewritten as 
\begin{align}
d_k &\ge  W^k_{\vect{x}_k}   \left\{ \frac{1}{k} \log \frac{W^k(\hat{Y}^k | \vect{x}_k)}{P_{\overline{Y}^k}(\hat{Y}^k)} > R + 2 \gamma \right\}  \label{eq:ineq13b}
\end{align}
and 
\begin{align}
\limsup_{k \rightarrow \infty} d_k \le \delta,~~~~~
\lim_{k \rightarrow \infty} P_{\vect{x}_k} = P_\gamma, \label{eq:ineq14b}
\end{align} 
respectively. The following lemma is of use.

\begin{e_lem} \label{lem:bad_dist}
Assume that $\vect{x}_k \in \mathcal{X}^k~(k=n_1, n_2, \cdots) $ satisfies \eqref{eq:ineq13b} and \eqref{eq:ineq14b} with some $\delta \in [0,1)$ and $P_\gamma \in \mathcal{P}(\mathcal{X})$, where $k=n_1, n_2, \cdots$ denotes either odd or even numbers.
If $k$ denotes odd numbers, then 
\begin{align}
P_\gamma \in A_{Y_1}(2 \gamma), \label{eq:good_dist2}
\end{align}
whereas if $k$ denotes even numbers, then
\begin{align}
P_\gamma \in A_{Y_2}(2 \gamma). \label{eq:good_dist3}
\end{align}
\end{e_lem}
\begin{e_proof}
Let $k=n_1, n_2, \cdots$ denote odd numbers. Suppose that $P_\gamma \not\in A_{Y_1}(2 \gamma)$.
From the right inequality in \eqref{eq:ineq14b} we obtain $P_{\vect{x}_k} \not\in A_{Y_1}(\gamma)$ for all large $k$.
Watanabe and Hayahi \cite[Lemma 2]{Watanabe-Hayashi2014} have shown that if $\vect{y} \in T_{W_1,\gamma}^k(\vect{x}_k)$, then
\begin{align}
\vect{y} \not\in T_{Y_1, \gamma'}^k.
\end{align}
Further, if $\vect{y} \not\in T_{Y_1, \gamma'}^k$, then $P_{\overline{Y}^k}(\vect{y}) =0$ by definition, and thus
\begin{align}
\frac{1}{k} \log \frac{W^k(\vect{y} | \vect{x}_k)}{P_{\overline{Y}^k}(\vect{y})} > R + 2 \gamma,
\end{align}
Therefore, for all $\vect{y} \in \mathcal{Y}^k$ we have
\begin{align}
\vect{1}\left\{ \vect{y} \in T_{W_1, \gamma}^k(\vect{x}_k) \right\}  &\le \vect{1}\left\{ \frac{1}{k} \log \frac{W^k(\vect{y} | \vect{x}_k)}{P_{\overline{Y}^k}(\vect{y})} > R + 2 \gamma \right\}. \nonumber
\end{align}
Since the set of $\gamma$-typical sequences $T_{W_1, \gamma}^k(\vect{x}_k)$ satisfies
\begin{align}
W^k_{\vect{x}_k}   \left\{ \hat{Y}^k \in T_{W_1, \gamma}^k(\vect{x}_k) \right\} \rightarrow 1 ~~~(k \rightarrow \infty),
\end{align}
this inequality and \eqref{eq:ineq13b} leads to 
\begin{align}
\lim_{k \rightarrow \infty} d_k = 1, \label{eq:bad_dist}
\end{align}
which is a contradiction, and hence \eqref{eq:good_dist2} holds. 

In the case of even numbers $k=n_1, n_2, \cdots$, \eqref{eq:good_dist3} can be proven analogously.
\QED
\end{e_proof}

Since $P_{\overline{Y}^k}(\vect{y}) \le P_{Y^k}(\vect{y})/ \tau_k$ for all $\vect{y} \in \mathcal{Y}^k$, we can bound the right-hand side of \eqref{eq:ineq13b} from below as
\begin{align}
d_k &\ge  W^k_{\vect{x}_k}   \left\{ \frac{1}{k} \log \frac{W^k(\hat{Y}^k | \vect{x}_k)}{P_{Y^k}(\hat{Y}^k)} > R + 2 \gamma + \frac{1}{k} \log \frac{1}{\tau_n} \right\}  \nonumber \\
&\ge  W^k_{\vect{x}_k}  \left\{ \frac{1}{k} \log \frac{W^k(\hat{Y}^k | \vect{x}_k)}{P_{Y^k}(\hat{Y}^k)} > R + 3 \gamma \right\} ~~~(k \ge k_0), \label{eq:ineq13c}
\end{align}
where the second inequality holds for all large odd numbers $k$.
Since the random variable
\begin{align}
\frac{1}{k} \log \frac{W^k(\hat{Y}^k | \vect{x}_k)}{P_{Y^k}(\hat{Y}^k)} = \frac{1}{k} \sum_{i=1}^k \log \frac{W_1(\hat{Y}_i | x_{k,i})}{P_{Y_1}(\hat{Y}_i)}
\end{align}
is a sum of \emph{conditionally} independent random variables given $\hat{X}^k = \vect{x}_k = (x_{k,1}, x_{k,2}, \ldots, x_{k,k})$, 
its expected value can be evaluated as
\begin{align}
& \E \left\{ \frac{1}{k} \log \frac{W^k(\hat{Y}^k | \vect{x}_k)}{P_{Y^k}(\hat{Y}^k)} \Big| \hat{X}^k = \vect{x}_k \right\} \nonumber \\
&~~=  \frac{1}{k}  \sum_{i=1}^k \sum_{b \in \mathcal{Y}} W_1(b|x_{k,i}) \log \frac{W_1(b | x_{k,i})}{P_{Y_1}(b)}  \nonumber \\
&~~=  \frac{1}{k}  \sum_{a \in \mathcal{X}} k P_{\vect{x}_k} (a)\sum_{b \in \mathcal{Y}} W_1(b|x_{k,i}) \log \frac{W_1(b | x_{k,i})}{P_{Y_1}(b)}  \nonumber \\
&~~=: D(W_1 || P_{Y_1} | P_{\vect{x}_k}), \label{eq:ineq15}
\end{align} 
 where $D(W_1 || P_{Y_1} | P_{\vect{x}_k})$ is the conditional divergence between $W_1$ and $P_{Y_1}$ given $P_{\vect{x}_k} \in \mathcal{P}(\mathcal{X})$.
Then, we can invoke the weak law of large numbers and under the conditional probability distribution $ W^k_{\vect{x}_k}$, yielding
\begin{align}
&\limsup_{k \rightarrow \infty}  W^k_{\vect{x}_k}  \left\{ \frac{1}{k} \log \frac{W^k(\hat{Y}^k | \vect{x}_k)}{P_{Y^k}(\hat{Y}^k)} > R + 3 \gamma \right\} \nonumber \\
&~~~~= \left\{
\begin{array}{ll}
 0 & \mathrm{if~}  D(W_1 || P_{Y_1} | P_\gamma) < R + 3 \gamma, \\
 1 & \mathrm{if~}   D(W_1 || P_{Y_1} | P_\gamma) > R + 3 \gamma
\end{array}
\right. 
\end{align}
and from the left inequality in \eqref{eq:ineq14b} and \eqref{eq:ineq13c},
we obtain 
\begin{align}
 R + 3 \gamma \ge   D(W_1 || P_{Y_1} | P_\gamma) . \label{eq:ineq16}
\end{align}
Since $\gamma >0 $ is arbitrary,
taking the limit $\gamma \downarrow 0$ for both sides, we obtain
\begin{align}
R &\ge \lim_{\gamma \downarrow 0}  D(W_1 || P_{Y_1} | P_\gamma)  \nonumber \\
 &=  D(W_1 || P_{\hat{Y}_1} | P_{\hat{X}_1})=  I(\hat{X}_1, \hat{Y}_1) \label{eq:ineq17}
\end{align}
with some $\hat{X}_1 \in B(X_1, W_1)$, where $\hat{Y}_1$ denotes the output via $W_1$ due to input $\hat{X}_1$.
Here, we have used the fact that $P_\gamma \in A_{Y_1}(\gamma) $ by Lemma \ref{lem:bad_dist} and $A_{Y_1}(\gamma) \rightarrow B(X_1, W_1)$ as $\gamma \downarrow 0$.
Thus, we have 
\begin{align}
R\ge  \inf_{\hat{X}_1 \in B (X_1, W_1)} I (\hat{X}_1; \hat{Y}_1), \label{eq:R_LB1}
\end{align}
completing the proof of \eqref{eq:converse_ineq1}.

(ii) 
To show \eqref{eq:converse_ineq2}, we first fix an arbitrary
\begin{align}
R > \inf_{\hat{\vect{X}} \in B_\delta^* (\vect{X}, \vect{W})} \sup_{\substack{\varepsilon \ge 0, \\ \vect{Z} \in \tilde{B}_{\varepsilon} (\hat{\vect{Y}})}}   \overline{D}_{\varepsilon} (\vect{W}|| \vect{Z} | \hat{\vect{X}}). \label{eq:rate_setting2}
\end{align}
Recall that we can write $X^n = X_{j(n)}^n$ and $W^n = W_{j(n)}^n$ and the corresponding output is $Y^n = Y_{j(n)}^n$ with definition \eqref{eq:channel_index}.
Let $\gamma > 0$ be arbitrarily fixed. We define $\gamma' := |\mathcal{X}| \gamma$, $\tau_n := \Pr \{ Y^n \in T_{Y_{j(n)}, \gamma'}^n \}$ and set $P_{\overline{Y}^n}$ again as in \eqref{eq:overlineY}.
Then, from the property of the set of $\gamma'$-typical sequences $T_{Y_{j(n)}, \gamma'}^n$, we have \eqref{eq:dist_convergence1}.

Now, for any general source $\hat{\vect{X}}=\{\hat{X}^n\}_{n=1}^\infty$ it is easily verified that
\begin{align}
\sup_{\substack{\varepsilon \ge 0, \\ \vect{Z} \in \tilde{B}_{\varepsilon} (\hat{\vect{Y}})}}   \overline{D}_{\varepsilon} (\vect{W}|| \vect{Z} | \hat{\vect{X}}) \ge \sup_{\substack{\varepsilon \ge 0, \\ \vect{Z} \in \tilde{B}_{\varepsilon}^* (\hat{\vect{Y}})}}   \overline{D}_{\varepsilon}^* (\vect{W}|| \vect{Z} | \hat{\vect{X}}),  \label{eq:D_relations}
\end{align}
where $\hat{\vect{Y}} =\{\hat{Y}^n\}_{n=1}^\infty$ denotes the output via $\vect{W}$ due to input $\hat{\vect{X}}$ and we define
\begin{align}
&\overline{D}_\varepsilon^* (\vect{W}|| \vect{Z}| \hat{\vect{X}}) :=  \mbox{\rm $\varepsilon$p$^*$-}\limsup_{n \rightarrow \infty} \frac{1}{n} \log \frac{W^n(\awidehat{Y}^n|\awidehat{X}^n)}{P_{Z^n}(\awidehat{Y}^n)}, \nonumber \\
&\tilde{B}_\varepsilon^* (\vect{Y}) := \left\{ \vect{Z} =\left\{ Z^n \right\}_{n=1}^\infty : \,  \liminf_{n \rightarrow \infty} d(P_{Y^n}, P_{Z^n}) \le \varepsilon \right\} \nonumber
\end{align}
since for all $\varepsilon > 0 $ and $\vect{Z}$ it holds that
\begin{align}
\overline{D}_{\varepsilon} (\vect{W}|| \vect{Z} | \hat{\vect{X}}) \ge \overline{D}_{\varepsilon}^* (\vect{W}|| \vect{Z} | \hat{\vect{X}}). 
\end{align}
We can see that by \eqref{eq:rate_setting2} and \eqref{eq:D_relations} there exists an $\hat{\vect{X}} \in B_\delta^* (\vect{X}, \vect{W})$ satisfying
\begin{align}
R &> \sup_{\substack{\varepsilon \ge 0, \\ \vect{Z} \in \tilde{B}_{\varepsilon}^* (\hat{\vect{Y}})}}   \overline{D}_{\varepsilon}^* (\vect{W}|| \vect{Z} | \hat{\vect{X}}) - \gamma \nonumber \\
 & \ge \overline{D}_{\delta}^* (\vect{W}|| \overline{\vect{Y}} | \hat{\vect{X}}) - \gamma \label{eq:LB3}
\end{align}
where to derive \eqref{eq:LB3} we have used that fact that $\overline{\vect{Y}} \in \tilde{B}_{\delta}^* (\hat{\vect{Y}})$ which follows from $\hat{\vect{X}} \in B_\delta^* (\vect{X}, \vect{W})$ and \eqref{eq:dist_convergence1} with the triangle inequality:
\begin{align}
&\liminf_{n \rightarrow \infty} d(P_{\hat{Y}^n}, P_{\overline{Y}^n}) \nonumber \\
&~~~\le \lim_{n \rightarrow \infty} d(P_{Y^n}, P_{\overline{Y}^n}) + \liminf_{n \rightarrow \infty} d(P_{Y^n}, P_{\hat{Y}^n}) \le \delta.
\end{align}
Equation \eqref{eq:LB3} implies that there exists some $\{ d_n > 0\}_{n = 1}^{\infty}$
satisfying
\begin{align}
\liminf_{n \rightarrow \infty} d_n \le \delta \label{eq:seq_dn1}
\end{align}
and
\begin{align}
d_n &\ge \Pr  \left\{ \frac{1}{n} \log \frac{W^n(\hat{Y}^n | \hat{X}^n)}{P_{\overline{Y}^n}(\hat{Y}^n)} > R + 2 \gamma \right\}  \label{eq:ineq21}
\end{align}
for all $n = 1, 2, \cdots$. Also, \eqref{eq:seq_dn1} indicates that at least one of the following inequalities holds:
\begin{align}
\liminf_{k \rightarrow \infty} d_{2k + 1} \le \delta~~~\mathrm{or}~~~\liminf_{k \rightarrow \infty} d_{2k} \le \delta. \label{eq:seq_dn2}
\end{align}

First, we assume that 
\begin{align}
\liminf_{\tilde{k} \rightarrow \infty} d_{\tilde{k}} \le \delta \label{eq:seq_dn4}
\end{align} 
for \emph{odd} $\tilde{k} = 1, 3, \cdots$. 
Similarly to the derivation of \eqref{eq:ineq13b} and \eqref{eq:ineq14b}, 
\eqref{eq:LB3} indicates that there exists some $\vect{x}_n \in \mathcal{X}^n$, a sequence $k = n_1, n_2, \cdots$, where $n_1, n_2, \cdots$ are \emph{odd numbers}, and $P_\gamma \in \mathcal{P}(\mathcal{X})$ such that 
\begin{align}
d_{k} &\ge  W^{k}_{\vect{x}_{k}}   \left\{ \frac{1}{k} \log \frac{W^{k}(\tilde{Y}^{k} | \vect{x}_k)}{P_{\overline{Y}^{k}}(\tilde{Y}^{k})} > R + 2 \gamma \right\}  . \label{eq:ineq33b}
\end{align}
and 
\begin{align}
\lim_{k \rightarrow \infty} d_{k} \le \delta,~~~~~
\lim_{k \rightarrow \infty} P_{\vect{x}_{k}} = P_\gamma. \label{eq:ineq34b}
\end{align}

From Lemma \ref{lem:bad_dist}, we have
\begin{align}
P_\gamma \in A_{Y_1}(2 \gamma). \label{eq:good_dist4}
\end{align}
Then, we can invoke the weak law of large numbers as in the derivation of \eqref{eq:R_LB1} to yield 
\begin{align}
R &\ge \inf_{\hat{X}_1 \in B (X_1, W_1)} I (\hat{X}_1; \hat{Y}_1) \nonumber \\
&\ge \inf_{\hat{X}_2 \in B (X_2, W_2)} I (\hat{X}_2; \hat{Y}_2),  \label{eq:ineq37}
\end{align}
where we have used \eqref{eq:ordered_system} for the last inequality.
Thus, we obtain \eqref{eq:converse_ineq2}.

In the case where \eqref{eq:seq_dn4} holds for \emph{even} $\tilde{k} = 2, 4, \cdots $, we can show \eqref{eq:ineq37} in the analogous way.
\QED


%

\end{document}